\documentclass[aps,a4paper,prl,twocolumn,superscriptaddress]{revtex4-2}
\usepackage{graphicx}
\usepackage[english]{babel}
\usepackage{amsmath}
\usepackage{amsfonts}
\usepackage{amssymb}
\usepackage{xcolor}
\usepackage{blindtext}
\usepackage{subfigure}%
\usepackage{hyperref}
\hypersetup{colorlinks,linkcolor=blue,urlcolor=blue,citecolor=blue}
\usepackage{braket}
\usepackage[super]{nth}
\usepackage{placeins}

\newcommand{\imp}{\mathrm{imp}}
\newcommand{\host}{\mathrm{host}}
\newcommand{\Ghost}{G^\host}
\newcommand{\Gimp}{G^\imp}
\newcommand{\DV}{\Delta V}

\newcommand{\vc}[1]{\ensuremath\boldsymbol{#1}}
\renewcommand{\vec}[1]{\vc{#1}}

\begin{document}

\title{High-throughput magnetic co-doping and design of exchange interactions in a topological insulator}

\author{Rubel Mozumder}
\affiliation{Peter Gr\"{u}nberg Institut (PGI-1), Forschungszentrum J\"{u}lich and JARA, 52425 J\"{u}lich, Germany}

\author{Johannes Wasmer}
\affiliation{Peter Gr\"{u}nberg Institut (PGI-1), Forschungszentrum J\"{u}lich and JARA, 52425 J\"{u}lich, Germany}
\affiliation{Institute for Theoretical Physics, RWTH Aachen University, 52074 Aachen, Germany}

\author{David Antognini Silva}
\affiliation{Peter Gr\"{u}nberg Institut (PGI-1), Forschungszentrum J\"{u}lich and JARA, 52425 J\"{u}lich, Germany}
\affiliation{Institute for Theoretical Physics, RWTH Aachen University, 52074 Aachen, Germany}

\author{Stefan Blügel}
\affiliation{Peter Gr\"{u}nberg Institut (PGI-1), Forschungszentrum J\"{u}lich and JARA, 52425 J\"{u}lich, Germany}

\author{Philipp R\"{u}{\ss}mann}
\email{p.ruessmann@fz-juelich.de}
\affiliation{Peter Gr\"{u}nberg Institut (PGI-1), Forschungszentrum J\"{u}lich and JARA, 52425 J\"{u}lich, Germany}
\affiliation{Institute of Theoretical Physics and Astrophysics, University of Würzburg, 97074 Würzburg, Germany}

\begin{abstract}

Using high-throughput automation of ab-initio impurity embedding simulations, we created a database of $3d$ and $4d$ transition metal defects embedded into the prototypical topological insulator (TI) Bi$_2$Te$_3$. We simulate both single impurities as well as impurity dimers at different impurity-impurity distances inside the topological insulator matrix. We extract changes to magnetic moments, analyze the polarizability of non-magnetic impurity atoms via nearby magnetic impurity atoms and calculate the exchange coupling constants for a Heisenberg Hamiltonian. 
We uncover chemical trends in the exchange coupling constants and discuss the impurities' potential with respect to magnetic order in the fields of quantum anomalous Hall insulators and topological quantum computing. In particular, we predict that co-doping of different magnetic dopants is a viable strategy to engineer the magnetic ground state in magnetic TIs.

\end{abstract}


\maketitle

\section{Introduction}

In recent years, Topological insulators (TIs) have emerged as a revolutionary class of materials, characterized by an insulating bulk and conducting surface states protected by time-reversal symmetry \cite{Hasan2010}. Breaking time-reversal symmetry in TI materials by magnetic doping turns them into so-called magnetic TIs (MTIs) \cite{Tokura2019, Bernevig2022}, where exotic quantum phenomena such as the quantum anomalous Hall (QAH) effect are found \cite{Chang2013, Chang2015, Kou2015, Bestwick2015,Grauer2015}, leading to the possibility of dissipationless transport by the edge states in the absence of a magnetic field \cite{QAHreview}. Pioneering experimental realization of the QAH effect in MTI were achieved by Cr-doped \cite{Chang2013} and V-doped \cite{Chang2015} $\mathrm{(Bi,Sb)_2}\mathrm{Te_3}$, where quantized transport was seen at temperatures ($T<100\mathrm{mK}$). The low critical temperature for the QAH effect remains an open challenge that is attacked by different strategies. On the one hand, new stochiometric materials like MnBi$_2$Te$_4$ \cite{Deng2020}, or transition-metal dichalcogenite van der Waals heterostructures are explored \cite{Li2021}. On the other hand, in MTIs,  modulation doping \cite{Mogi2015} or co-doping strategies \cite{Ou2018} are viable options to raise the critical temperature of QAH materials.

Aside from the QAH phase, MTIs have great potential for spintronics \cite{Fan2014}, and offer the possibility of realizing the axion insulator state that, in contrast to the QAH phase, relies on antiferromagnetic order \cite{QAHreview, Wang2015, Zhuo2023}. Furthermore, in the emerging field of topological superconductivity, combining topological electronic structures with magnetic doping and superconductivity is a pathway towards topological superconducting materials with possibilities for fault-tolerant quantum computing architectures \cite{Tokura2019}.

Thus, controlling the magnetic properties of MTIs, e.g., the size of the exchange splitting or the long-range magnetic ordering of the magnetic atoms, poses a materials optimization challenge that is addressed in this paper. Based on high-throughput ab-initio impurity embedding calculations \cite{Ruessmann2020}, we simulate transition metal defects and dimers of impurities embedded into the prototypical TI Bi$_2$Te$_3$, thus extending the JuDiT database (\href{https://go.fzj.de/judit}{https://go.fzj.de/judit}) of calculated impurity properties embedded into topological insulators \cite{Ruessmann2020} with co-doping of impurity dimers. Using density functional theory (DFT) simulations, we calculate their magnetic properties, the exchange coupling constant at different impurity distances that map the magnetic interactions onto a classical Heisenberg Hamiltonian, and discuss the influence of co-doping to the magnetic order. Overall, our database includes several thousand impurity dimer calculations.

The paper is organized as follows. In Sec.~\ref{seq:methods} the computational methodology is introduced. Then, in Sec.~\ref{seq:results}, the results of the impurity database of this work is presented and discussed. Finally, Sec.~\ref{seq:conclusion} concludes the paper with a summary and an outlook.

\section{Methods}
\label{seq:methods}

\subsection{Ab-initio impurity embedding}

\begin{figure}
    \centering
    \includegraphics[width=0.95\linewidth]{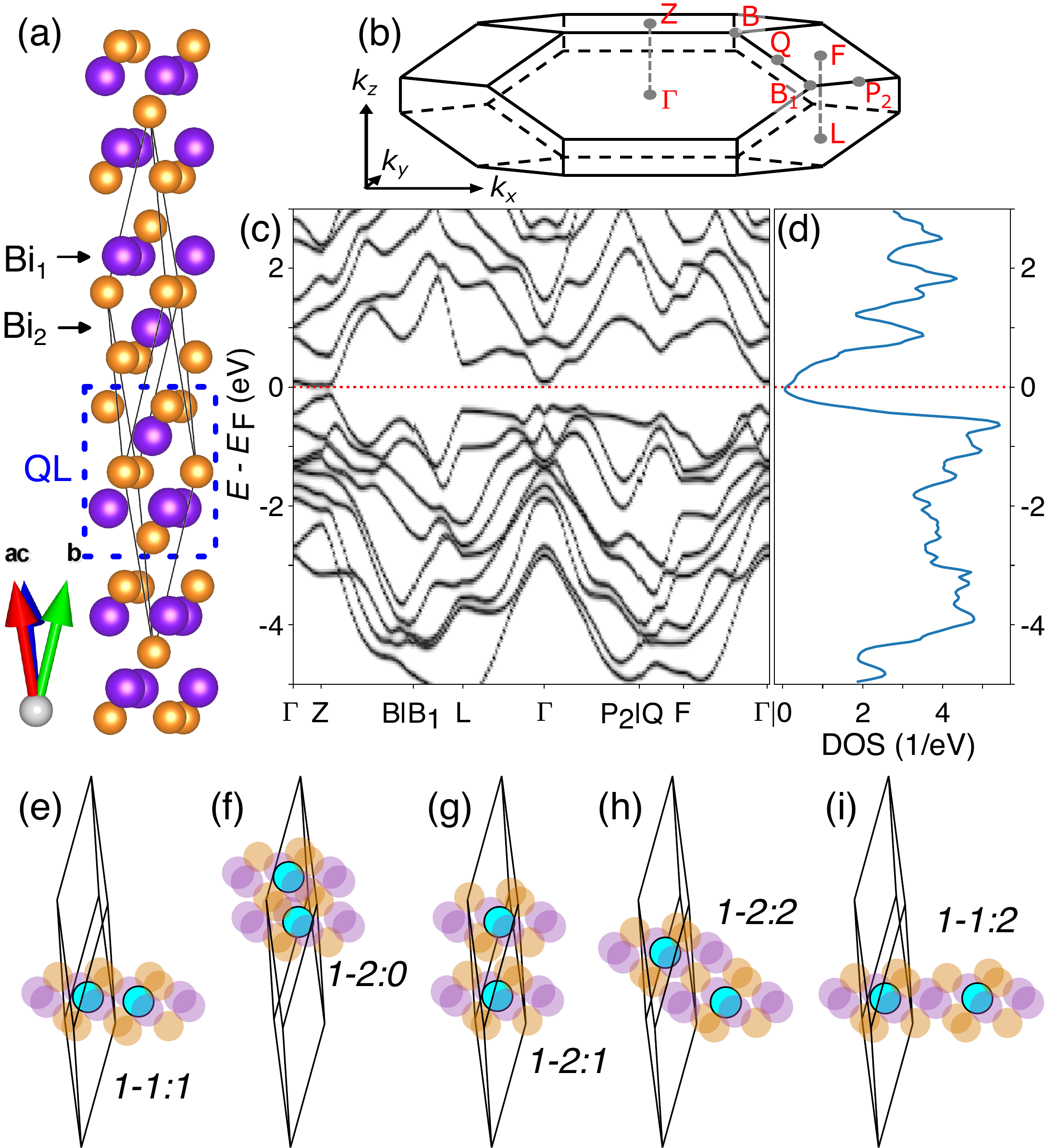}
    \caption{Calculation setup. (a) Crystal structure of the Bi$_2$Te$_3$ host crystal where the quintuple layer (QL) structure with the two Bi sites Bi$_1$, Bi$_2$ is highlighted. (c) Electronic band structure along paths connecting the high-symmetry points indicated in the Brillouin zone in (b) and (d) corresponding density of states. (e-i) Double-impurity clusters for increasing distances between the impurities 
    ($R=4.38\mathrm{\AA}$, $4.79\mathrm{\AA}$, $6.10\mathrm{\AA}$, $6.49\mathrm{\AA}$, $7.59\mathrm{\AA}$).
    The impurity atoms (blue) are substituting Bi$_{1/2}$ as indicated by the labels of the form ``$\mu-\nu:o$'' where $\mu,\nu$ indicate the Bi layer and $o$ indexes the shell of nearest neighbor distance.}
    \label{fig:setup}
\end{figure}

The DFT results of this work are produced within the local density approximation (LDA) \cite{Vosko1980} using the full-potential relativistic Korringa-Kohn-Rostoker Green's function method (KKR) \cite{Ebert2011} as implemented in the {\tt JuKKR} code package \cite{jukkr}. We use the experimental crystal structure of bulk Bi$_2$Te$_3$ \cite{ExpLatticeBi2Te3} shown in Fig.~\ref{fig:setup}(a). Throughout this work we use an $\ell_{max} = 3$ cutoff in the angular momentum expansion with an exact description of the atomic cells \cite{Stefanou1990,Stefanou1991}, and we use an energy contour in the integration of the charge density with 63 complex-valued energy points and a $40^3$ $k$-points in the Brillouin zone integration of the self-consistent calculations. The error in the charge density integration arising from the finite $\ell_\mathrm{max}$ cutoff is corrected using Lloyd's formula~\cite{Zeller2004}. This leads to a bulk insulating Bi$_2$Te$_3$ host crystal as shown in the electronic band structure and corresponding DOS in Fig.~\ref{fig:setup}(b-d). 
All our calculations include the effect of spin-orbit coupling (SOC) self-consistently, which is essential to reproduce the inverted (i.e., topological) band structure of Bi$_2$Te$_3$ around the $\Gamma$ point.

The Green function formulation of the KKR method allows to include impurities efficiently into crystalline solids \cite{Ebert2011,Bauer2013} which, in this work, is done for $3d$ and $4d$ transition metal impurities substituting Bi sites in the Bi$_2$Te$_3$ host crystal structure as shown in Fig.~\ref{fig:setup}. Here we make use of the Dyson equation
\begin{equation}
    \Gimp = \Ghost + \Ghost \DV \Gimp
    \label{eq:Dyson}
\end{equation}
where $\Ghost$ is the Green function of the crystalline host system, 
$\DV = V^{\imp}-V^{\host}$ is the difference in the potential introduced due to the presence of the impurity and $\Gimp$ is the Green function that describes the impurity embedded into the periodic host crystal. Importantly, the change in the potential, $\DV$, occurs only in a small region around the impurity. Thus, the Dyson equation can be solved in a small real space area around the impurity site called the \emph{impurity cluster}. Having neighboring host atoms included in the impurity cluster is, however, required for a proper charge screening of the embedded impurity. In our work the impurity cluster contains all atoms within a cutoff radius of 4.5\AA, i.e., including the first three neighboring shells of host atoms. For the impurity embedding step and the subsequent calculation of the $J_{ij}$ parameters (cf.~Eq.~\eqref{eq:Heisenberg}), we increase the $k$-point integration grid in the calculation of $\Ghost$ for Eq.~\eqref{eq:Dyson} to $100^3$ $k$-points.

\subsection{Extended Heisenberg Hamiltonian}

The focus of this work is the evaluation of the magnetic properties of $3d$ and $4d$ transition metal impurities in Bi$_2$Te$_3$. Therefore our impurity embedding focuses on dimers of impurity atoms at different distances, illustrated in Fig.~\ref{fig:setup}(e-f). These DFT calculations for impurity dimers are then mapped onto the Heisenberg Hamiltonian to describe the magnetic interactions among  impurities in the form
\begin{equation}
    H = -\frac{1}{2}\sum_{i,j}J_{ij} \, \vec{S}_i \cdot \vec{S}_j 
    \label{eq:Heisenberg}
\end{equation}
Here, $i,j$ label the magnetic atoms at different distances corresponding to the connecting vectors $\vec{R}_{ij}=\vec{R}_j-\vec{R}_i$, $\vec{S}_i={\vec{m}_i}/{|\vec{m}_i|}$ denotes the unit vector indicating the direction of the magnetic moment $\vec{m}_i$ at site $i$, and $J_{ij}$ are the isotropic exchange interaction that are responsible for the long-range magnetic ordering discussed in Sec.~\ref{seq:Tc}. The Green function formulation of KKR allows to calculate the exchange parameters directly from the DFT electronic structure of impurity dimers using the method of infinitesimal rotations \cite{Liechtenstein1987}. In this work we are modelling random distributions of impurity atoms within the TI host crystal at low impurity concentrations that neglect structural changes to the host crystal. This random orientation will lead to an averaging-out of directional Dzyaloshinskii-Moriya interaction (DMI) that would otherwise require to include additional terms of the form $\vec{D}_{ij}\cdot(\vec{S}_i\times\vec{S}_j)$ in the Heisenberg Hamiltonian and which favor a canting of spins and non-collinear spin textures. Since we assume that the DMI will average out over the random impurity configurations, we focus our analysis on changes of the impurities' magnetic moments and on their $J_{ij}$ parameters.

\subsection{High-throughput workflow}

We consider all 20 impurity dimers of $3d$ and $4d$ impurities up to a distance of 12.5\AA\ between the two impurities. Figure~\ref{fig:setup}(e-i) and Fig.~\ref{figa:alldists} in the Supplemental Material illustrate the geometry of the impurity dimers we include in this study. For each impurity dimer at a given distance, $20\times21 / 2 = 210$ unique impurity dimers need to be calculated self-consistently to take into account possible charge transfer between the impurities that can influence the impurities' magnetic properties. This results in a large number of self-consistent DFT calculations for dimers at different distances and the subsequent extraction of $J_{ij}$ parameters. At larger distances we restrict our calculations to pairs of impurites that are magnetic in the single impurity limit (i.e., where $|m|>10^{-3}\mu_B$). Overall we calculate more than 2,000 unique impurity dimers and extract their exchange coupling parameters.

\begin{figure}
    \centering
    \includegraphics[width=\linewidth, trim={0 0.2cm 0 0},clip]{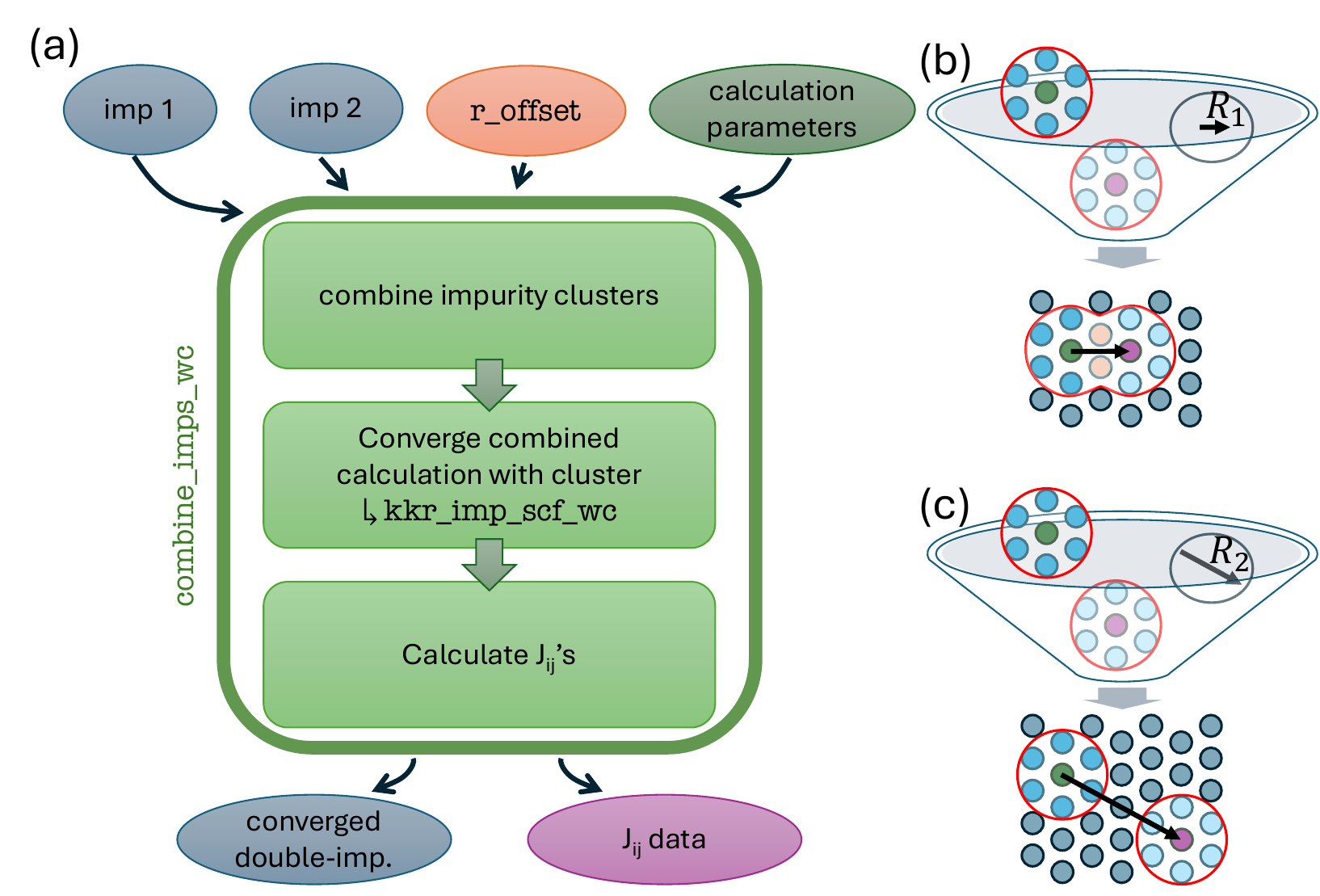}
    \caption{AiiDA workflow to combine impurity calculations to larger clusters. (a) Outline of the {\tt combine\_imps\_wc} workchain of the AiiDA-KKR plugin. Based on two impurities as input together with a definition of the offset vector between the impurities ({\tt r\_offset}) the workflow first combines the impurity clusters to one larger impurity cluster that is embedded into the host crystal in a self consistent calculation making use of the {\tt kkr\_imp\_scf\_wc} workchain before finally the $J_{ij}$ parameters are extracted in a post-processing step. (b,c) Illustration of two inputs to the {\tt combine\_imps\_wc} workchain for different offset vectors $\vec{R}$. For small $\vec{R}$ the single impurity embedding clusters might overlap where duplicate positions (orange sites in (b)) are removed before the self-consistent DFT calculation for the dimer is pursued.}
    \label{fig:app2}
\end{figure}

The large number of DFT calculations in this study are therefore conveniently orchestrated using the {AiiDA-KKR} \cite{aiida-kkr-paper, aiida-kkr-code} plugin to the AiiDA high-throughput infrastructure \cite{aiida} where we used the \texttt{submission\_controller} of AiiDA-JuTools in the workload management \cite{jutools}. For this purpose we have extended the capabilities of AiiDA-KKR with a new workflow called {\tt combine\_imps\_wc}, which is sketched in Fig.~\ref{fig:app2}. This workflow is capable of starting from two impurity calculations, combines them into a larger impurity cluster, converges the DFT calculation for this combined cluster, and, finally, extracts the $J_{ij}$ parameters.  The {\tt combine\_imps\_wc} workflow is also capable of extending an impurity cluster consisting of a dimer or more impurity atoms and add another impurity cluster. In this way small nanostructures of impurities can be constructed and embedded self-consistently into the host crystal.

\section{Results and discussion}
\label{seq:results}

\subsection{Single impurity properties}

\begin{figure}
    \centering
    \includegraphics[width=\linewidth]{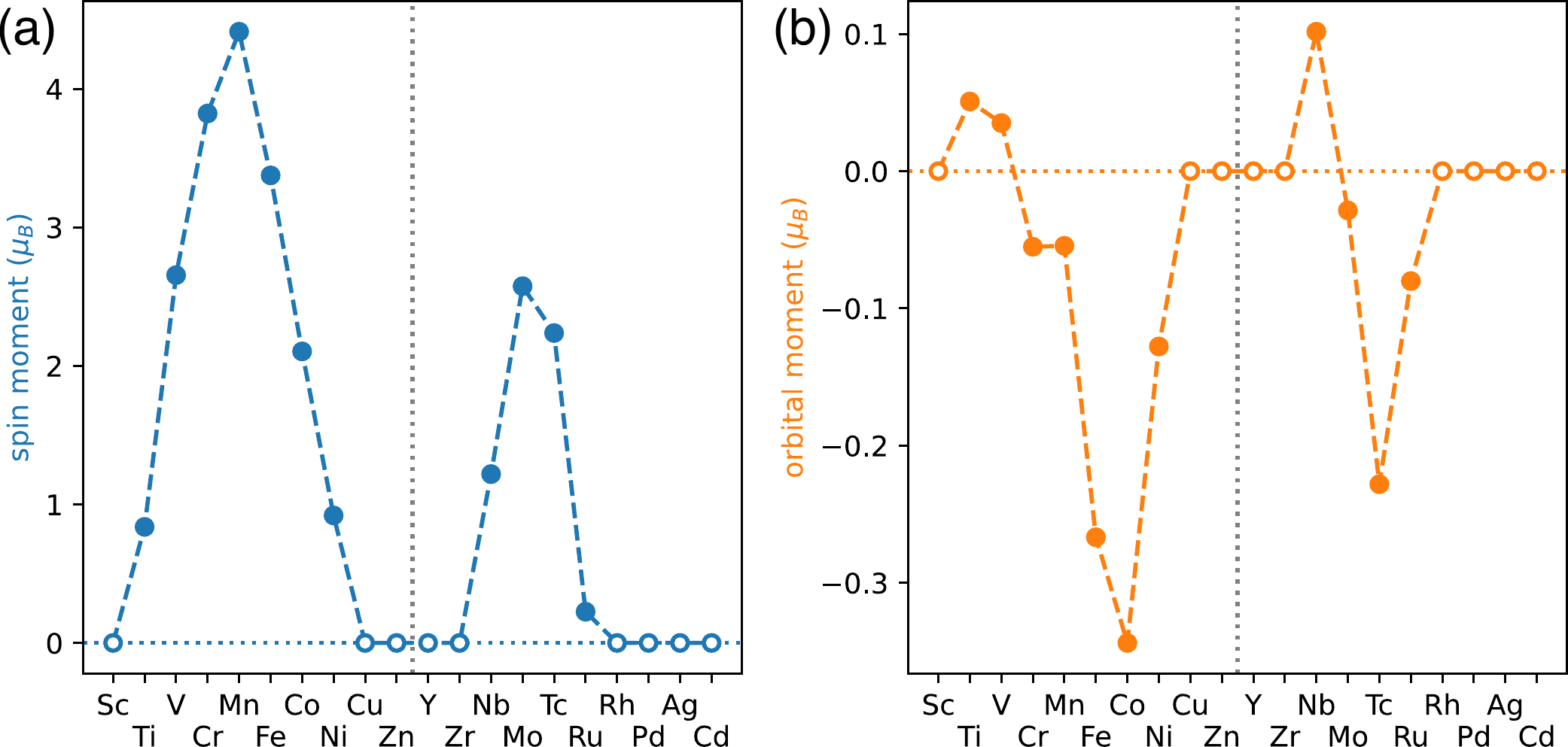}
    \caption{Single impurity magnetic (a) spin- and (b) orbital moments for $3d$- and $4d$-impurities embedded as substitutional defects on Bi sites in Bi$_2$Te$_3$.}
    \label{fig:single}
\end{figure}

Figure~\ref{fig:single} summarize the magnetic moments of single impurities embedded as subsitutional defects on Bi sites within bulk Bi$_2$Te$_3$. Going through the $3d$ and $4d$ series of transition metal impurities, we find that the $3d$ impurities from V to Co and the $4d$ impurities from Nb to Ru have a magnetic ground state where the magnetic moments reach a maximum following the well-known trends with increasing occupation of the $d$-shell of the impurities. The orbital moment generally follows Hund's rules with a sign change from positive to negative orbital moment for increasing $d$-shell filling.

\begin{figure}
    \centering
    \includegraphics[width=\linewidth]{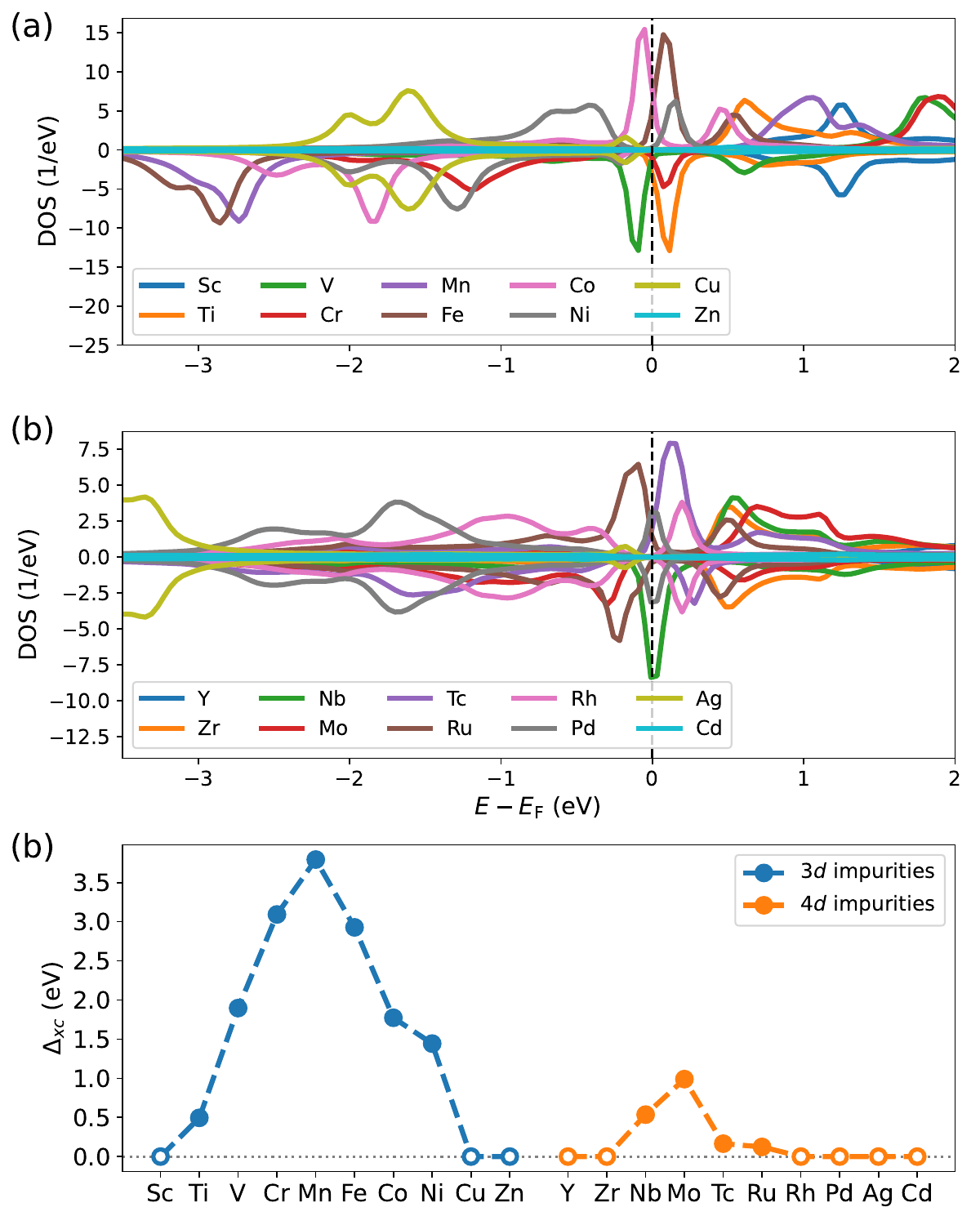}
    \caption{Density of states of single substitutional $M_\mathrm{Bi}$ defects for (a) $M=3d$ and (b) $M=4d$ impurities. The minority (majority) spin channel is shown with a positive (negative) sign of the DOS. (c) Size of the exchange splitting $\Delta_{xc}$ extracted from the energy difference between the maxima of minority and majority spin channels.}
    \label{fig:DOS}
\end{figure}

Figure~\ref{fig:DOS} summarizes the single impurity density of states for all twenty $3d$ and $4d$ defects in Bi$_2$Te$_3$. 
Some impurities show resonances of the $d$ states around the Fermi energy like V, Cr, Fe and Co from the $3d$ series and Nb, Tc or Pd from the $4d$ series. 
This is in line with earlier calculations for impurities in Sb$_2$Te$_3$ and it can be important in the materials optimization procedure when designer properties (such as a low DOS at the Fermi energy) are desirable~\cite{aiida-kkr-paper}.
From the difference of the position of the dominant peak in the minority and majority spin channels of the DOS, we estimate the size of the exchange splitting $\Delta_{xc}$ of the different impurities. This is summarized in Fig.~\ref{fig:DOS}(c) where we recover the same trend of $\Delta_{xc}$ with the $d$-shell as in the size of the spin moments (\textit{cf.}\ Fig.~\ref{fig:magmom}). The largest $\Delta_{xc}$ is found for Mn where the spin moment is also maximal. Within the series of $4d$ series of impurity elements the magnitude of the exchange splitting is smaller. Thus, using $4d$ dopants such as Mo can combine a low DOS at the Fermi level with a sizable spin moment and a moderate exchange splitting. This can be beneficial in the materials optimization challenge around magnetic TIs proximitized by a superconductor in the field of topological quantum computing~\cite{DiMiceli2023, Legendre2024}.  

\subsection{Co-doping-induced changes of the magnetic moments and polarizability}

\begin{figure}
    \centering
    \includegraphics[width=0.85\linewidth]{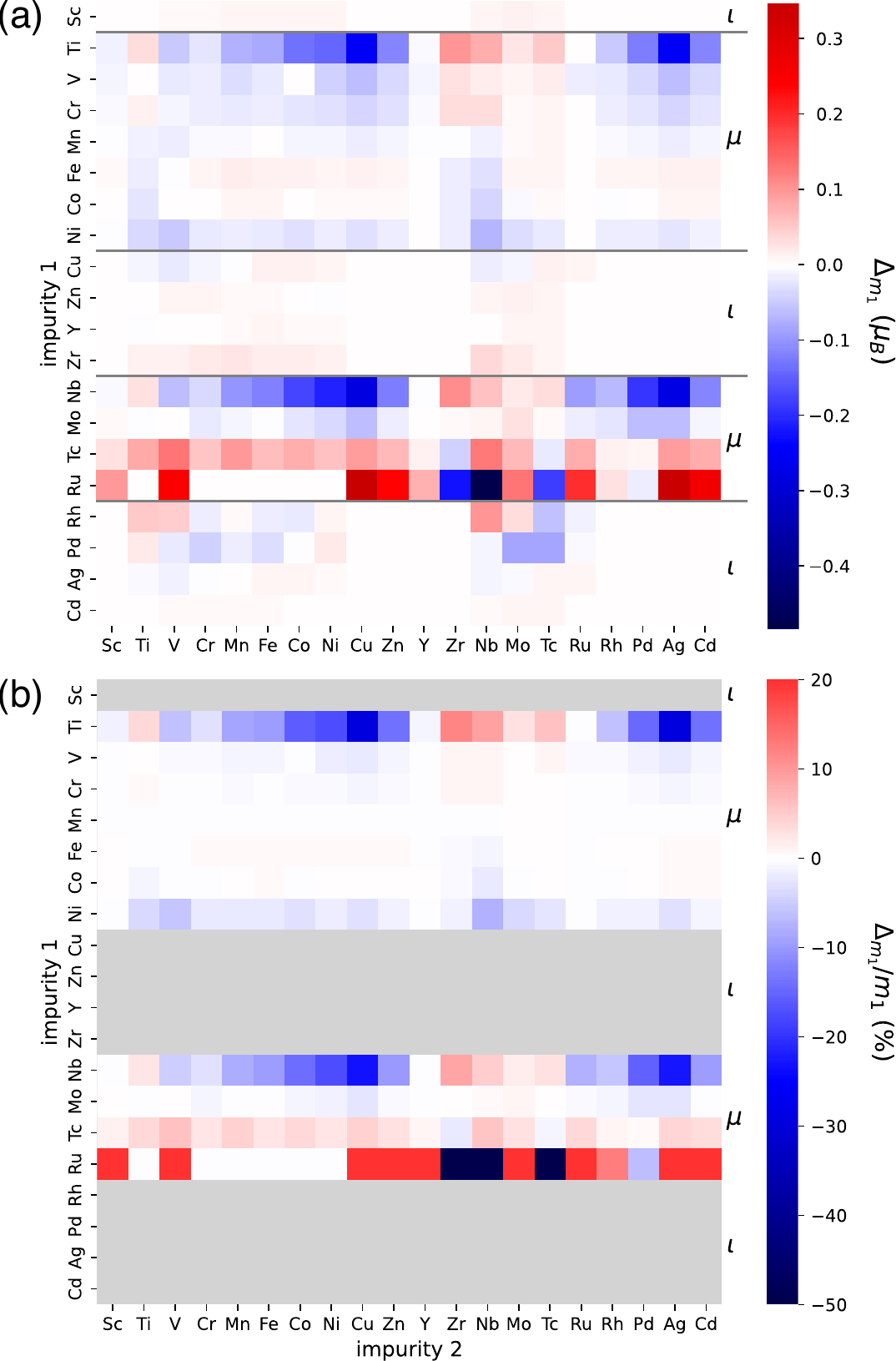}
    \caption{Changes in the magnetic moments for nearest-neighbor impurities ($R=4.38$\AA). (a) Absolute change $\Delta_{m_1}$ in $\mu_B$ of the spin moment of impurity 1 upon adding impurity 2 and (b) corresponding relative change in \%. The grey horizontal lines and the symbols $\iota$ ($\mu$) label non-magnetic (magnetic) impurities in the isolated single impurity case meaning that the magnetic moment change given for $\iota$ impurities denote induced moments due to nearest-neighbor magnetic atoms.}
    \label{fig:magmom}
\end{figure}

The effect of co-doping on the magnetic moments of impurities is shown in Fig.~\ref{fig:magmom} where the absolute and relative change in the spin magnetic moment for nearest neighbor impurities is shown in panel (a,b), respectively. We find changes to the magnetic moments of impurity 1 $\Delta m_1$ due to the presence of impurity 2 that reach $0.5\mu_B$ in magnitude. Magnetic impurities (marked by the symbol $\mu$ in Fig.~\ref{fig:magmom}) show larger changes to their magnetic moments where the largest changes are seen for impurities with weak magnetic moments of isolated single impurities like Ti ($m=0.84\mu_B$, $\max|\Delta m_1|=0.25\mu_B$), Nb ($1.22\mu_B$, $0.28\mu_B$), or Ru ($0.23\mu_B$, $0.48\mu_B$) \footnote{Note that for Ru $\max|\Delta m_1| > m$ because for Ru--Nb the Ru moment increases in magnitude but also flips the sign to an antiparallel alignment with respect to the Nb moment.}. Magnetic impurities from the middle of the $3d$ series that show large magnetic moments (V to Co) are only weakly affected as seen from the small relative change of $\leq 2\%$ in the magnetic moments shown in Fig.~\ref{fig:magmom}(b). Furthermore, similarities between the Ti and Nb suggest that a similar filling of the valence $d$-shell can explain the chemical trends which we observe.  

Initially non-magnetic impurities (marked by the symbol $\iota$ in Fig.~\ref{fig:magmom}) can be polarized by the presence of nearby magnetic defects which is particularly pronounced for Rh and Pd with induced moments reaching magnitudes of $0.1\mu_B$ for nearby Nb, Mo or Tc atoms. A positive (negative) sign of the induced moment then indicates parallel (anti-parallel) alignment of the spins in the two impurity atoms. The high polarizability correlates with a peak in the DOS close to the Fermi energy as shown by the pink (grey) curves in Fig.~\ref{fig:DOS}(b) for Rh (Pd).

\subsection{Distance-dependent exchange coupling constants}

\begin{figure*}
    \centering
    \includegraphics[width=0.99\linewidth]{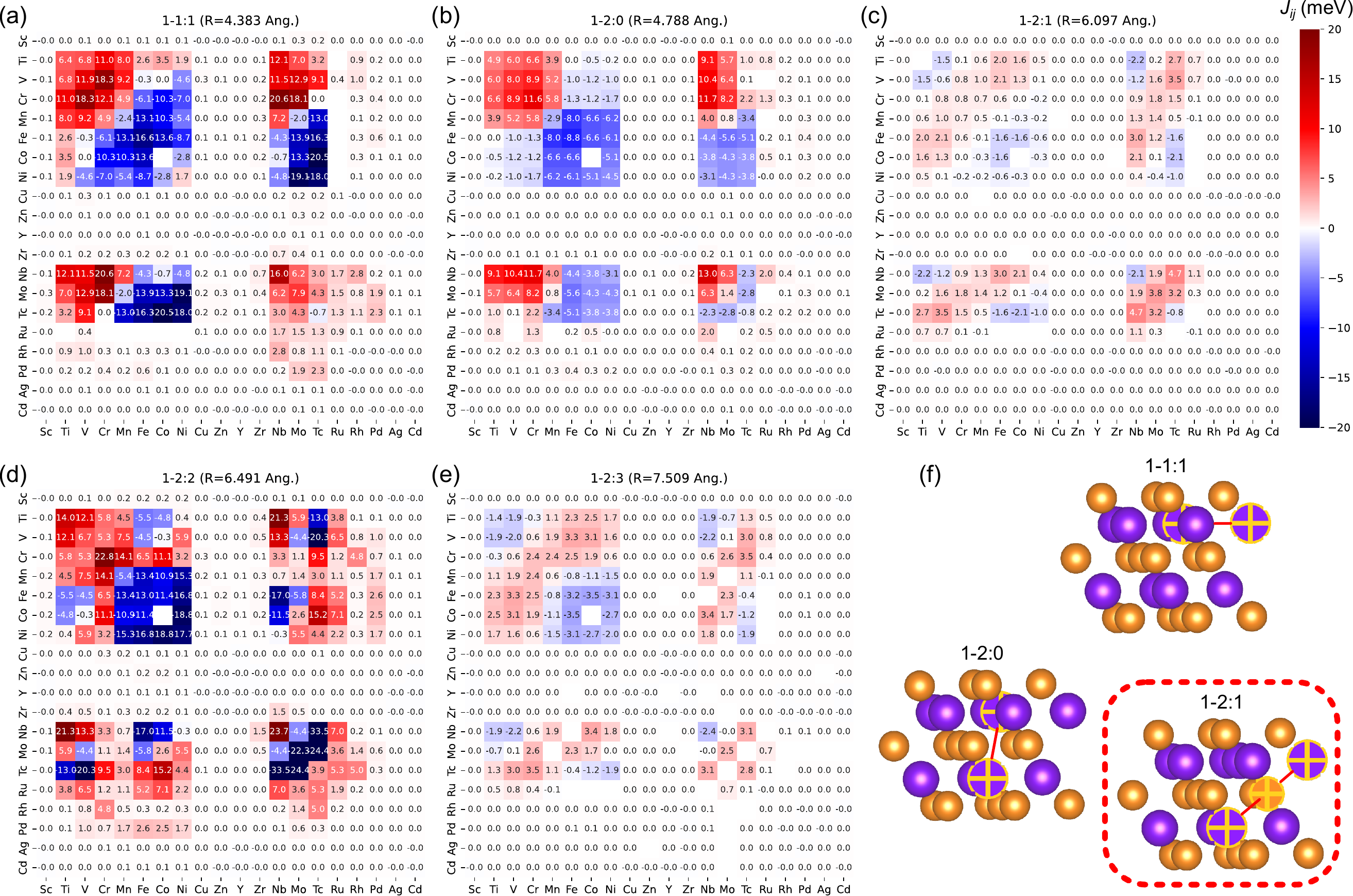}
    \caption{Pairwise Heisenberg exchange coupling parameters $J_{ij}$ for (a-e) the first five impurity-impurity distances where the impurities are from the $3d$ ($Z_\mathrm{imp}=21-30$, i.e., Sc to Zn) and $4d$ ($Z_\mathrm{imp}=39-48$, i.e., Y to Cd) series. (f) Visualization of the three impurity dimer configurations that lead to the largest exchange coupling constants (shown in a,b,d) which are among subsitutional Bi sites (labeled Bi$_1$ and Bi$_2$, cf.~Fig.~\ref{fig:setup}) within the same Bi$_2$Te$_3$ quintuple layer. The highlighted configuration (1-2:1) shows the largest magnitude of the $J_{ij}$ values due to the linear Bi$_1$-Te-Bi$_2$ geometry maximising the electronic interactions between impurities at these sites.}
    \label{fig:Jijs}
\end{figure*}

Figure~\ref{fig:Jijs} shows maps of the $J_{ij}$ couplings for all impurity dimers of $3d$ and $4d$ dopants at the first five distances between $4.4$\AA\ and $7.5$\AA. 
As expected, dimers with non-magnetic impurities (e.g., Sc, Cu, Zn, Y, Zr, Rh, Pd, Ag, Cd) lead to vanishing exchange couplings which serves as a consistency check for our method. 
Overall, we find clear chemical trends following the $d$-orbital filling of the magnetic dopants. This can be seen, for instance, in the red to blue color change that indicates a sign change when following the diagonal within the block of magnetic $3d$ impurities (i.e., from Ti to Ni) in panels (a,b). However, at larger distances we find two sign changes from $J_{ij}<0$ to $J_{ij}>0$ and back to $J_{ij}<0$ as seen in panels (c,e). Similar trends are visible for dimers within the $4d$ block of dopants, where, however, fewer impurities are magnetic to begin with. Also co-doping of $3d$ with $4d$ impurities follow the this trend. 

With increasing distances, the magnitude of the $J_{ij}$'s generally decay but then suddenly recover to large values of $|J_{ij}|\geq20$meV again at a distance of $6.5$\AA\ shown in Fig.~\ref{fig:Jijs}(d). A close inspection of the impurity dimer in real space for the \nth{4} neighbor reveals its special character. As illustrated in Fig.~\ref{fig:Jijs}(f), the subsitutional Bi sites of this dimer lie on a straight line with an additional Te atom in between. This special geometry can lead to the Kramers-Anderson superexchange mechanism where the exchange interaction is mediated by virtual electron hoppings from the first impurity to the nonmagnetic Te site and then onward to the second impurity \cite{Koch2012}. Following the Goodenough-Kanamori rule, this superexchange mechanism is maximal for 180$^\circ$ orientations between the subsitutional impurities on Bi sites in the Bi$_1-$Te$-$Bi$_2$ configuration. 

A closer analysis reveals an oscillating sign of the exchange interactions as seen, for instance, for Ti--Ti dimers with two sign changes from positive to negative (and vice versa) within the first five neighbors. This can be attributed to a competition of different exchange mechanisms including direct exchange of overlapping impurity states at short distances, superexchange for the \nth{4} neighbor, and RKKY-like oscillatory behavior for larger distances \cite{Peixoto2020}.

\begin{figure}
    \centering
    \includegraphics[width=\linewidth, trim={0 0.5cm 1cm 0cm},clip]{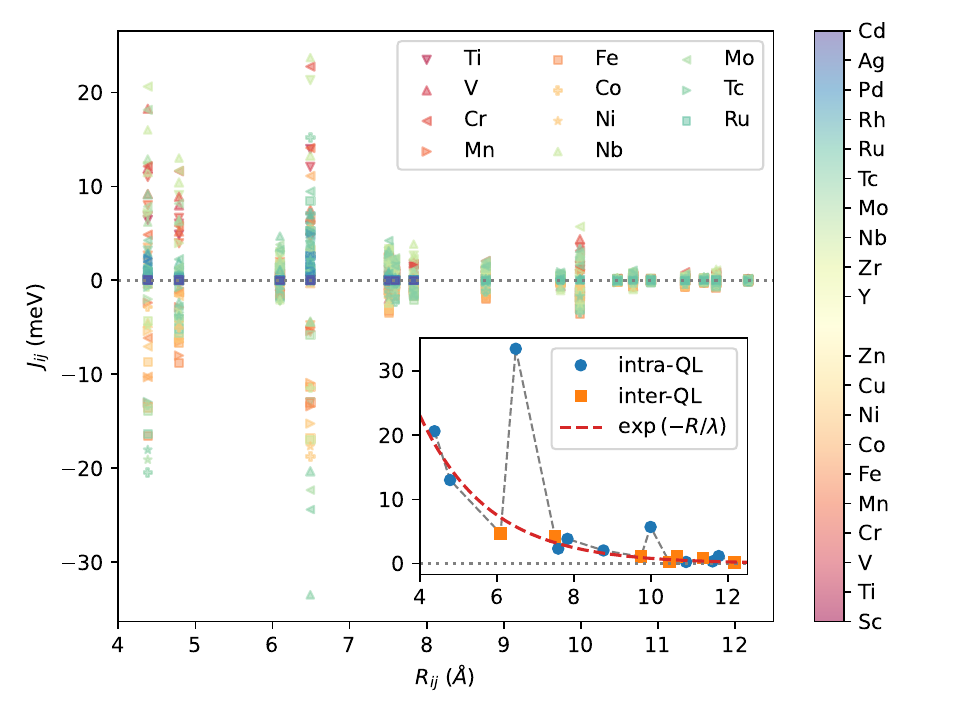}
    \caption{Decay of the Heisenberg exchange coupling constants with distance. The symbols indicate the type of atom $i$ and the colors the type of atom $j$ of the respective $J_{ij}$ pairs. The inset shows the maximal absolute value for each shell where the color of the symbol indicates couplings within the same quintuple layer (``intra-QL'', circles) or from one to the next (``inter-QL'', squares) quintuple layer above or below. The red dashed line in the inset shows a fit for an exponential decay (neglecting the outlier at the maximal value at $R_{ij}=6.49\mathrm{\AA}$) giving a decay constant of $\lambda=1.79\mathrm{\AA}$.}
    \label{fig:JijsRadial}
\end{figure}

Figure~\ref{fig:JijsRadial} shows the distance dependence of the $J_{ij}$'s for all combinations of magnetic impurities included in our study. We find both positive and negative values for the exchange coupling depending on the type and the distance between impurities in the dimers. Generally, there is an overall decay of the magnitude of $J_{ij}$ with distance as shown in the inset of Fig.~\ref{fig:JijsRadial} where the maximal absolute value of the $J_{ij}$ at a given distance between the impurities is shown. We find a striking outlier to this general trend for the \nth{4} neighbor ($R_{ij}=6.5\mathrm{\AA}$) where the largest overall values are found despite the large distance between the impurities due to the 180$^\circ$ alignment discussed above. 
Neglecting the outlier of the \nth{4} neighbor, we can characterize the decay of the magnitude of the $J_{ij}$'s well with an exponential function of the form $J(R) = J_0\cdot e^{-R/\lambda}$ with $J_0=215.2$meV and $\lambda=1.55$\AA.

\subsection{Long-range magnetic ordering}
\label{seq:Tc}

\begin{figure}
    \centering
    \includegraphics[width=\linewidth]{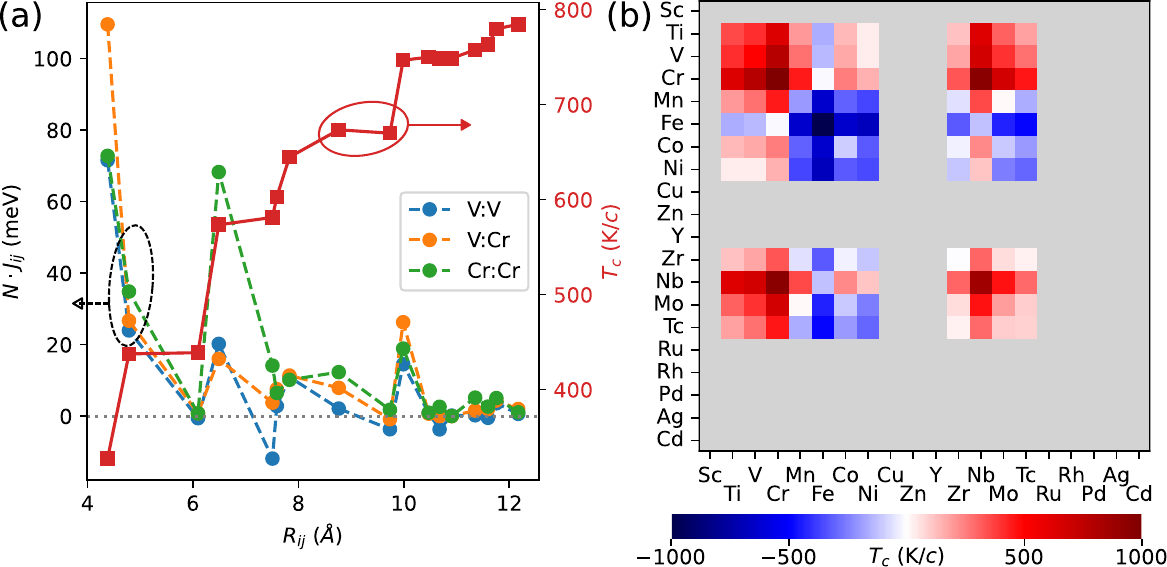}
    \caption{Tendency for long-range magnetic order estimated by the mean-field ordering temperature. (a) Convergence of the mean-field critical temperature (Eq.~\eqref{eq:Tc}) as function of the distance between two V and/or Cr impurities leading to $T_c= 784.5 \mathrm{K}/c$ (assuming equal impurity concentrations $c=c_i=c_j$ for V and Cr). Broken lines report the pairwise $J_{ij}$'s multiplied by the number of atoms per shell and full lines indicate the $T_c$ as function of the distance between the impurity atoms. (b) Map of the resulting estimate for the mean-field $T_c$ for all impurity combinations assuming equal concentration of both impurity types ($c=c_i=c_j$) as described in the main text. Note that a positive (negative) $T_c$ indicates the tendency for ferromagnetic (antiferromagnetic) long-range ordering. Grey areas indicate non-magnetic impurities at site $i$ and/or $j$ leading a vanishing set of $J_{ij}$'s.}
    \label{fig:Tc}
\end{figure}

Finally, as a way to estimate the collective magnetic ordering for a given impurity configuration, we compute the mean-field critical temperature \cite{Pajda2001}
\begin{equation}
    T_c=\frac{1}{3}k_B\sum_{j>0} c_j J_{0,j}
    \label{eq:Tc}
\end{equation}
where the sum over the pairs $j$ counts all equivalent pairs within shells at constant $R_{ij}$ and $c_{j}$ are the concentrations of magnetic impurities at site $j$ (i.e., with magnetic moments $m_{i,j}\geq 10^{-3}\mu_B$) which are considered up to distances of $R_{ij}\leq 12.5\mathrm{\AA}$ \footnote{Note that the prefactor here is $1/3$ instead of $2/3$ due to the different definition of the Heisenberg Hamiltonian Eq.~\eqref{eq:Heisenberg} that contains a factor $1/2$ in our work.}. Figure~\ref{fig:Tc} shows the convergence of the $T_c$ together with the contributions of the $J_{ij}$'s arising from the different shells for V--Cr co-doping in Bi$_2$Te$_3$. The overall exponential decay of the $J_{ij}$'s counteracts the increasing number of neighbors in the shells at greater distances which leads to a slow convergence of the critical temperature with distance between impurities. It is therefore essential to include large distances of $R_{ij}\geq10\mathrm{\AA}$ in this analysis.

For simplicity, we assume equal concentration of the two impurity kinds at sites $i$ and $j$ in the calculation of the $T_c$. Therefore, we simply average the $T_c$ arising from the different impurity-impurity configurations in the case of co-doping of two kinds of impurities. For instance, to describe V--Cr alloys we calculate $T_c= (T_c[\mathrm{V-V}]+T_c[\mathrm{Cr-Cr}]+T_c[\mathrm{V-Cr}])/3$. Figure~\ref{fig:Tc}(b) shows the resulting map of the $T_c$'s where the blue color indicates negative $T_c$ (antiferromagnetic ordering) while red colors indicate positive $T_c$ (ferromagnetic). 

Co-doping can be a viable strategy to increase the stability of a desired magnetic order as demonstrated by the Cr--V co-doped configuration. This is evident from the contributions to the converged $T_c$ for V--Cr pairs which show larger couplings for nearest neighbors as well as for the tenth neighbor which would be important at large (small) impurity concentrations when nearest (far away) neighbors contribute more to the magnetic ordering due to the difference in the average impurity distance with concentration. 

Overall, we find the largest positive $T_c$ values for Cr impurities which is also a well-known QAH material \cite{Chang2013, QAHreview}. Typical V- and Cr-doped TIs show magnetic ordering temperatures between $10-40\mathrm{K}$, depending on the concentration of magnetic atoms ($M$) which typically lies in the range of $c=10-20\%$ for ($M_c$Bi$_{1-c}$)$_2$Te$_3$ \cite{Kou2014, Checkelsky2014, QAHreview}. Using these concentrations we arrive at an estimate mean-field ordering temperature of $T_c=50-200\mathrm{K}$. The overestimation of the $T_c$ compared to experimental observations is a known deficiency of the mean-field approximation which, especially in the limit of dilute magnetic dopants, exaggerates the contribution of nearest neighbor impurity dimers that only occur seldom at low concentrations. In contrast, larger distances between impurities become more important where the pronounced local maxima at the \nth{4} and \nth{10} neighbor would stabilize more long-range magnetic order and lead to lower percolation thresholds and a more stable long-range magnetic order. All these effects are not included in the crude mean-field approximation we apply here which only serves as a rough estimate allowing us to judge the general tendency for magnetic ordering in these materials.

The tendency to overestimate the influence of short impurity distances is seen in a quantitative comparison of the mean-field estimate for the $T_c$ when the nearest neighbor contributions are neglected \footnote{In contrast to Eq.\eqref{eq:Tc} the summation starts at $j=2$ instead of $j=1$.}
\begin{equation}
    T_c' = \frac{1}{3}k_B \sum_{j>1} c J_{0,j}.
    \label{eq:Tcnofirst}
\end{equation}
For Cr, this leads to a moderate improvement in the estimated ordering temperature from $T_c/T_c^\mathrm{exp}=5.5$ to $T_c'/T_c^\mathrm{exp}=4$ with $T_c^\mathrm{exp}=40\mathrm{K}$ for an impurity concentration of $c=22\%$ \cite{Checkelsky2014}. For Mn impurities, our estimation from the mean-field approach even predict antiferromagnetic ordering which disagrees with the small ferromagnetic ordering temperature of $10\mathrm{K}$ at $c=9\%$ found in experiments \cite{Hor2010}. We speculate that this discrepancy could be related to different trends in the inter-QL vs. intra-QL exchange parameters (\textit{cf.}~Fig.~\ref{figa:Jijzdist}) where we find a sign change for Mn. As shown in the supplemental material in Fig.~\ref{figa:Tcinterintra}, the summation of the inter-QL $J_{ij}$'s results in a ferromagnetic ordering of the Mn atoms from different QLs, while the intra-QL ordering favors antiferromagnetic alignment within the same layer.  Finally, for V dopants at $c=12\%$ impurity concentration, we find values of $T_c' \approx 28\mathrm{K}$ with Eq.~\eqref{eq:Tcnofirst}, which almost exactly reproduces the experimental value of $30\mathrm{K}$ \cite{Kou2014} compared to the overestimation by a factor $T_c/T_c^\mathrm{exp}\approx 2$ when using the full summation in Eq.~\eqref{eq:Tc}.
Thus, a proper treatment of the magnetic ordering would include all these effects of the impurity distribution, e.g., in form of Monte-Carlo simulations for large supercells that explicitly take into account the percolation threshold of the different impurities at varying concentrations. However, this would go beyond the scope of our present study.

In summary, in our estimation for the magnetic order we find that co-doping of Cr with Ti, V, Mo or Nb or doping purely with $4d$ impurities like Nb and Mo can also lead to stable ferromagnetic ordering as indicated by large values of the mean-field $T_c$. In contrast, configurations with Mn, Fe, Co, or Ni impurities generally lead to antiferromagnetic order.

\section{Conclusion}
\label{seq:conclusion}

In this work have presented a computational database of impurity dimers embedded into the topological insulator Bi$_2$Te$_3$ based on high-throughput density-functional theory calculations. We have analyzed the impurities' DOS and exchange splitting and find a high magnetic polarizability by adjacent magnetic atoms of some transition metal impurities (Rh, Pd), which are non-magnetic as single defects. 
Our calculations of the Heisenberg exchange coupling constants reveal the chemical trends with regard to the impurities' $d$-shell filling. We find an overall exponential decay of the $J_{ij}$'s with distance between the impurities with notable exceptions at special distances where the $J_{ij}$'s are significantly larger in magnitude. This is traced back to the geometry of the TI's host crystal structure.

From the trends in the long-range magnetic ordering, we conclude that co-doping of Cr with other magnetic impurities like Ti, V, Mo or Nb is a viable strategy to strengthen ferromagnetic ordering, which might lead to more stable QAH materials.
Other impurity configurations with Mn or Fe generally favor antiferromagnetic interactions, that could be useful in the context of axion insulators. 
In the field of topological quantum computing, a different design strategy in MTIs might be to useful. In contrast to the QAH field where the most stable ferromagnetic ordering is sought-after, a smaller exchange splitting or (partial) antiferromagnetic couplings that could effectively weaken the pair-breaking potential for Cooper pairs in the proximity to a superconductor (SC) might be beneficial to optimize the proximity effect in MTI/SC interfaces.
While the material trends we find could serve as guide to future experiments, an experimental realization might be more complex. For instance, co-doping of Cr and V in Sb$_2$Te$_3$ can lead to the formation of undesired Cr$_2$Te$_3$ instead of the random alloying which we assume here \cite{Duffy2017}. Therefore additional studies are necessary in the active field of MTI materials.

\begin{acknowledgments}
\section*{Acknowledgements}

We thank Prof.\ Phivos Mavropoulos for valuable discussions. We acknowledge financial support of the Bavarian Ministry of Economic Affairs, Regional Development and Energy within the High-Tech Agenda Project ``Bausteine für das Quantencomputing auf Basis topologischer Materialien mit experimentellen und theoretischen Ansätzen''.
This work was performed as part of the Helmholtz School for Data Science in Life, Earth and Energy (HDS-LEE) and received funding from the Deutsche Forschungsgemeinschaft (DFG, German Research Foundation) under Germany's Excellence Strategy -- Cluster of Excellence Matter and Light for Quantum Computing (ML4Q) EXC 2004/1 -- 390534769, and from  the Helmholtz Association of German Research Centres, and from the European
Joint Virtual Lab on Artificial Intelligence, Data Analytics and Scalable Simulation (AIDAS). 
The authors are, furthermore, grateful for the computing time granted through JARA on the supercomputer JURECA~\cite{jureca} at Forschungszentrum Jülich (project id ``superint'') as well as computing time granted by the JARA Vergabegremium and provided on the JARA Partition part of the supercomputer CLAIX at RWTH Aachen University (project id ``jara0191'').

\end{acknowledgments}

\subsection*{Data and code availability}

The complete dataset for this paper, that includes the full data provenance of all calculations and the scripts used in the data analysis, is made publicly available in the materials cloud repository \cite{doi-dataset}. The source codes of the \texttt{JuKKR} code, and the AiiDA-KKR and AiiDA-JuTools plugins are available as open-source repositories from Refs.~\cite{jukkr2022, aiida-kkr-code, jutools}.

\subsection*{Author contributions}

RM, DAS and PR developed the \texttt{combine\_imps\_wc} workchain used for the impurity dimer calculations in this work. RM, JW, and PR analyzed the data. PR and SB designed the study and PR wrote the manuscript with the help of all co-authors.

\section*{Supplemental Material}
\setcounter{figure}{0}
\renewcommand{\thefigure}{A\arabic{figure}}

\begin{figure*}
    \centering
    \includegraphics[width=0.8\linewidth]{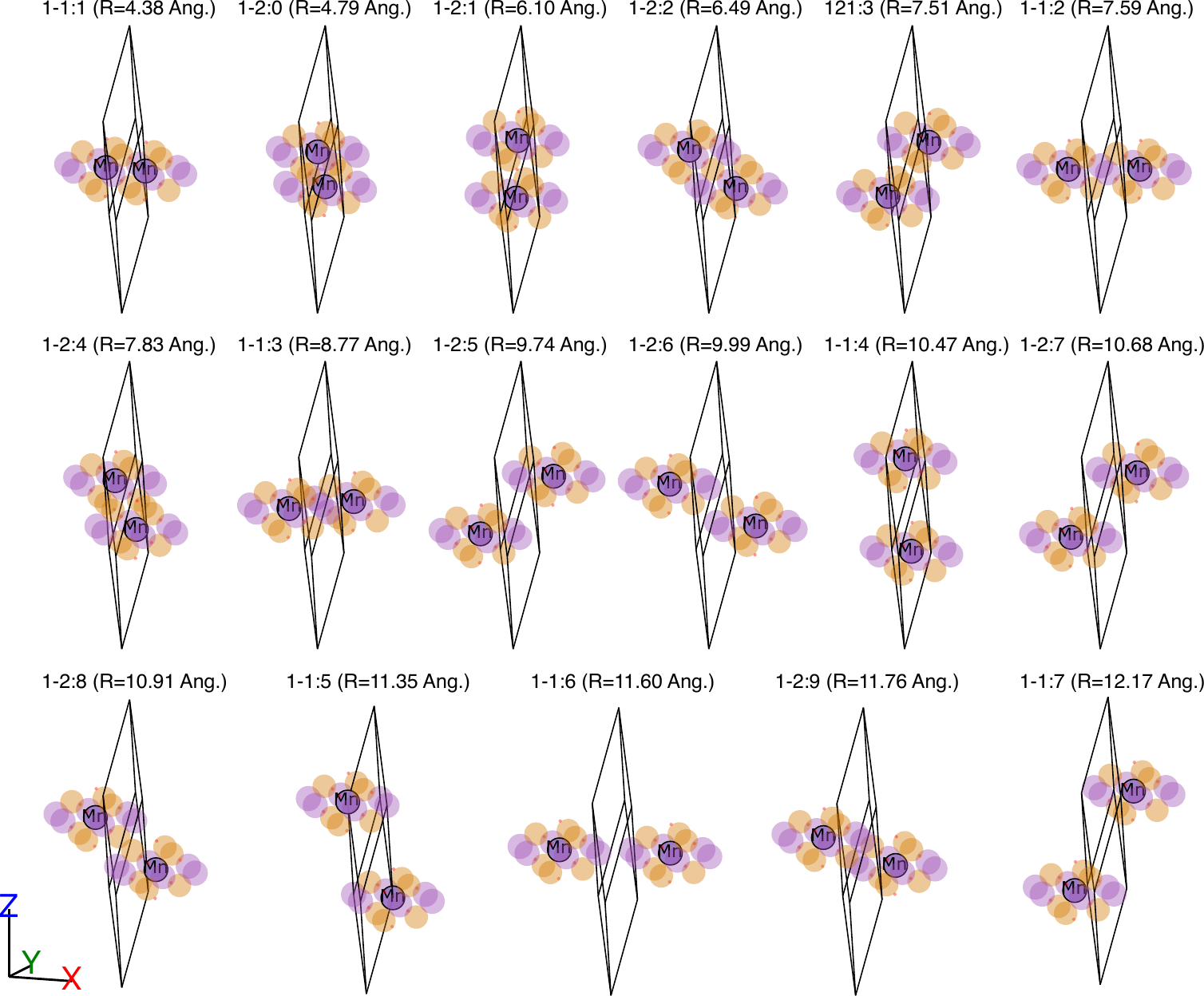}
    \caption{Impurity clusters for all 17 neighbors considered in this study. Shown are impurity configurations with Mn-Mn dimers for illustration. Note that host atoms from the Bi$_2$Te$_3$ matrix that are not part of the impurity cluster are not shown but enter the calculation via the $\Ghost$ in Eq.~\eqref{eq:Dyson} from the main text as boundary conditions in the impurity embedding step.}
    \label{figa:alldists}
\end{figure*}

\begin{figure*}
    \centering
    \includegraphics[width=\linewidth]{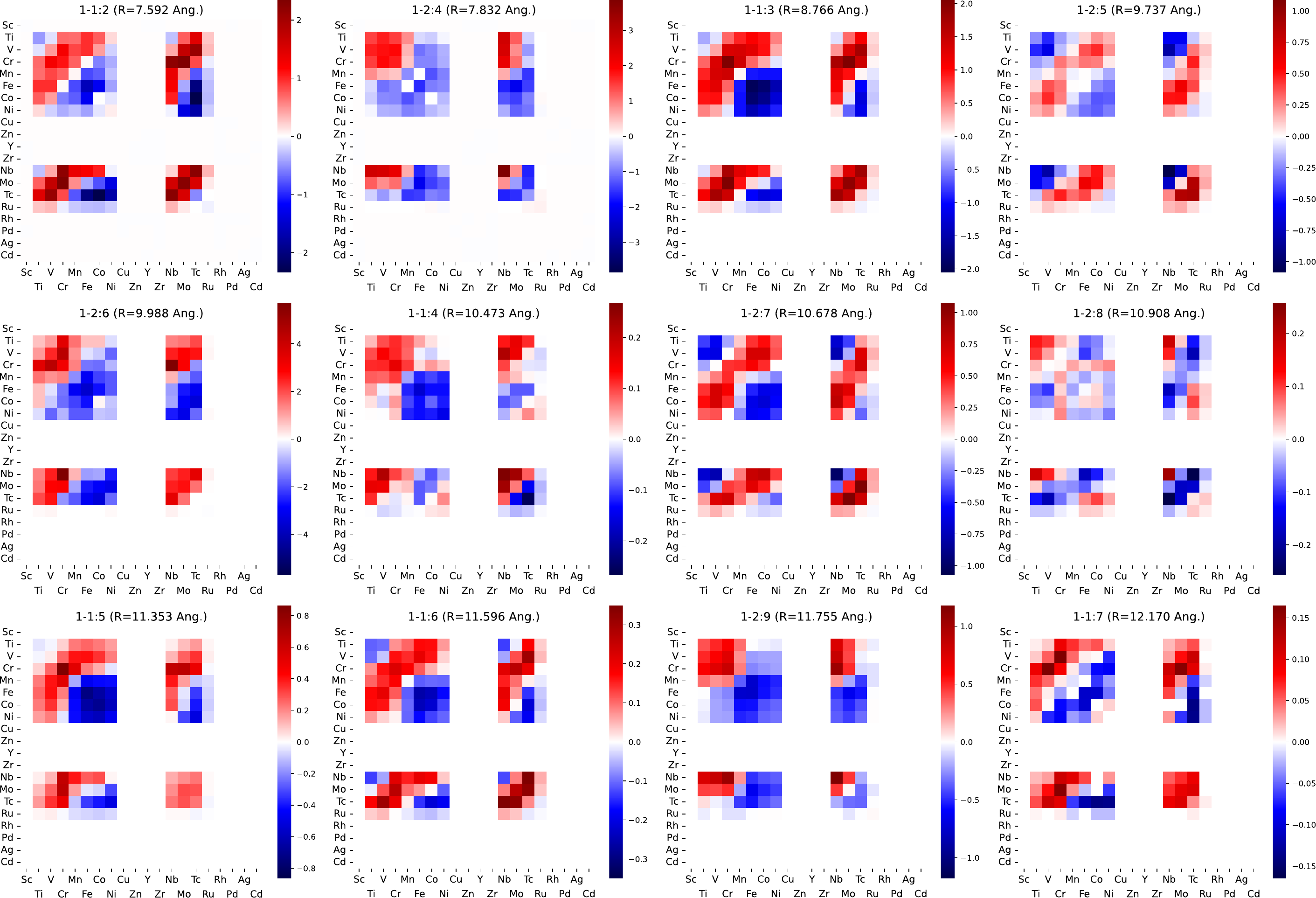}
    \caption{Maps of the $J_{ij}$ parameters (in meV) for all distances $R_{ij}\geq7.59$\AA\ (shorter distances are shown in Fig.\ref{fig:Jijs} of the main text). The title of each subplot indicates the impurity configuration and distance as introduced in Fig.~\ref{fig:setup} of the main text.}
    \label{figa:Jijall}
\end{figure*}

\begin{figure*}
    \centering
    \includegraphics[width=\linewidth]{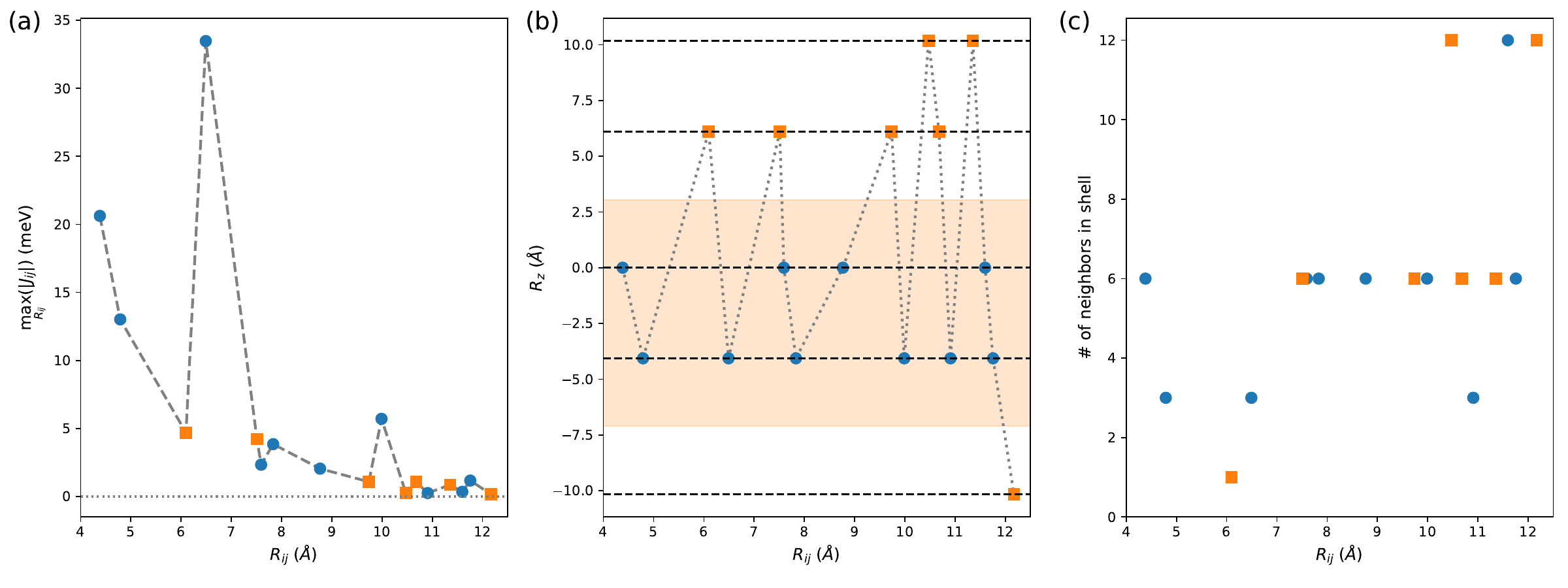}
    \caption{(a) Decay of intra-QL (blue circles) and inter-QL (orange squares) contributions to the $J_{ij}$'s with distance between the impurities. (b) $z$-component of the connecting vector $\vec{R}_{ij}$ highlighting coupling within the same QL (intra-QL) or from one QL to the neighboring QLs (inter-QL). (c) Number of equivalent neighbors in a shell at given distance $R_{ij}$.}
    \label{figa:Jijzdist}
\end{figure*}

\begin{figure*}
    \centering
    \includegraphics[width=\linewidth]{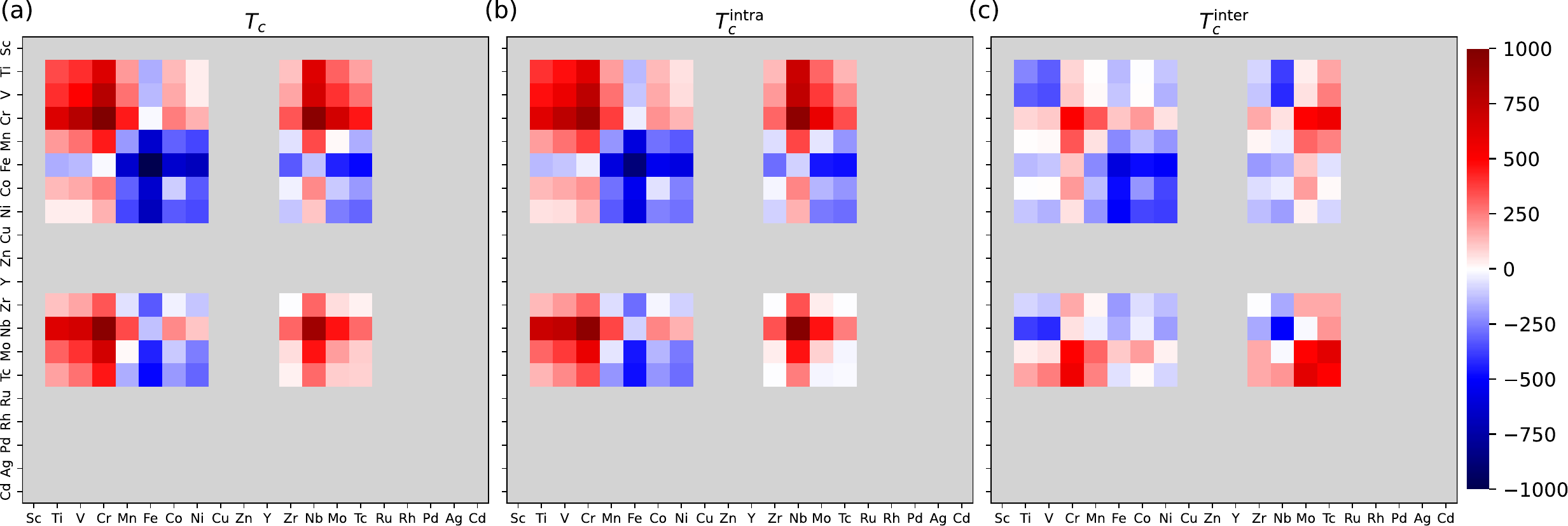}
    \caption{Mean field critical temperature (in $\mathrm{K}/c$, where $c$ is the impurity concentration) summed over (a) all $J_{ij}$'s, (b) only over the intra-QL $J_{ij}$'s and (c) summed only over the inter-QL $J_{ij}$'s (cf.~Fig.~\ref{figa:Jijzdist}).}
    \label{figa:Tcinterintra}
\end{figure*}

\bibliography{references.bib}

\end{document}